\documentclass[reprint,showpacs,amsmath,amssymb]{revtex4-1}

 \usepackage[colorlinks=true, linkcolor=blue, urlcolor=blue, citecolor=blue, anchorcolor=blue]{hyperref}
\usepackage{graphicx,amsmath,amssymb,bm,multirow}
\usepackage{amsthm}
\usepackage{amsfonts}
\usepackage{dcolumn}
\usepackage{bm}
\usepackage{mathrsfs}
\usepackage[usenames,dvipsnames]{color}
\newcommand{\bsub}{  \begin{subequations}}
\newcommand{\esub}{ \end{subequations}}

\usepackage[T1]{fontenc}
\usepackage[utf8]{inputenc}
\usepackage{graphicx,color,rotating,pifont}
\usepackage{mathtools}
\usepackage{siunitx} 
\usepackage{bm}
\usepackage{ae}

\usepackage{siunitx}
\usepackage{dcolumn}
\usepackage{txfonts}
\usepackage{tensor}
\usepackage{braket}
\usepackage{booktabs}
\usepackage{xspace}
\usepackage{mathrsfs}
\usepackage{amssymb} 
\usepackage{amsmath}
\usepackage{esvect}
\usepackage{diagbox}

\usepackage{hyperref}

\DeclareSIUnit{\fm}{\femto\meter}

\newcommand{\nuc}[2]{\ensuremath{{}^{#2}\mathrm{#1}}}

\newcommand{\beq}{\begin{equation}}
\newcommand{\eeq}{\end{equation}}
\newcommand{\beqn}{\begin{eqnarray}}
\newcommand{\eeqn}{\end{eqnarray}}

\begin{document}
\title{Multireference covariant density functional theory for shape coexistence and isomerism in $^{43}$S}

 \author{E. F. Zhou}
  \affiliation{School of Physics and Astronomy, Sun Yat-sen University, Zhuhai 519082, P. R. China}    

\author{X. Y. Wu} 
\email{xywu@jxnu.edu.cn}
\address{College of Physics and Communication Electronics, Jiangxi Normal University, Nanchang 330022, P. R. China}
\affiliation{Jiangxi Provincial Key Laboratory of Advanced Electronic Materials and Devices, Nanchang 330022, P. R. China}  

\author{J. Xiang} 
\email{jxiang@cqnu.edu.cn}
 \affiliation{College of Physics and Electronic Engineering, Chongqing Normal University, Chongqing 401331, P. R. China
 }

  \author{J. M. Yao}   
  \email{yaojm8@sysu.edu.cn} 
  \affiliation{School of Physics and Astronomy, Sun Yat-sen University, Zhuhai 519082, P. R. China} \affiliation{Guangdong Provincial Key Laboratory of Quantum Metrology and Sensing, Sun Yat-Sen University, Zhuhai 519082, P. R. China }

  \author{P. Ring}    
  \email{peter.ring@tum.de}
  \affiliation{ Physik Department, Technische Universität München, D-85748 Garching, Germany}

\date{\today}

\begin{abstract} 

We extend the multireference covariant density functional theory (MR-CDFT) to describe the low-lying states of the odd-mass nucleus $^{43}$S near the neutron magic number $N=28$ with shape coexistence. The wave functions of the low-lying states are constructed as superpositions of configurations with different intrinsic shapes and $K$ quantum numbers, projected onto good particle numbers and angular momenta. The MR-CDFT successfully reproduces the main features of the low-energy structure in $^{43}$S. Our results indicate that the ground state, $3/2^-_1$, is predominantly composed of the intruder prolate one-quasiparticle (1qp) configuration $\nu1/2^-[321]$. In contrast, the $7/2^-_1$ state is identified as a high-$K$ isomer, primarily built on the prolate 1qp configuration $\nu7/2^-[303]$. Additionally, the $3/2^-_2$ state is found to be an admixture dominated by an oblate configuration with $K^\pi = 1/2^-$, along with a small contribution from a prolate configuration with $K^\pi = 3/2^-$. These results demonstrate the capability of MR-CDFT to capture the intricate interplay among shape coexistence, $K$-mixing, and isomerism in the low-energy structure of odd-mass nuclei around $N = 28$.

\end{abstract}

\pacs{21.10.-k, 21.60.Jz, 21.10.Re}
\maketitle

 \section{Introduction }

The development of radioactive  ion beam facilities~\cite{Yano:2007,Glasmacher:2017,Ye:2025}  has significantly advanced nuclear physics research, enabling studies of nuclei far from the $\beta$-stability line, where the evolution of nuclear shell structure and the emergence of exotic excitation modes have garnered considerable attention~\cite{Caurier:2004RMP,Sorlin:2008PPNP,Otsuka:2018RMP,Tsunoda:2020natrue}.  A striking example is the evolution of the $N=28$ shell gap, a magic number that arises from strong spin-orbit coupling in the single-nucleon potential~\cite{Mayer:1949,Haxel:1949}, which drives the $f_{7/2}$ orbital significantly lower than the $p_{3/2}$ orbital. Experimental data reveal a gradual weakening of the $N=28$ shell gap in isotones lighter than \nuc{Ca}{48}. For instance, measurements of the beta-decay half-lives  of \nuc{S}{44} and \nuc{Cl}{45} revealed deviations from shell model predictions based on spherical configurations~\cite{Sorlin:1993S44half-life}, indicating the weakening of the $N=28$ shell effect. Subsequent Coulomb excitation experiments observed low  excitation energies of the  $2^+_1$ states and enhanced electric quadrupole transition strengths $B(E2; 0^+_1 \to 2^+_1)$ in \nuc{S}{40,42,44}~\cite{Scheit:1996S4042,Glasmacher:1997S44}.  Similar behavior has been reported in neighboring nuclei, such as \nuc{Si}{42}~\cite{Gade:2019Si42,Bastin:2007Si42,Takeuchi:2012Si42} and \nuc{Mg}{40}~\cite{Crawford:2019Mg40}. These observations indicate the onset of strong quadrupole collectivity in  neutron-rich $N=28$ isotones with proton number $Z<18$, leading to the crossing of the $\nu1/2^-[321]$ component of the $2\nu p_{3/2}$ orbital with the $\nu7/2^-[303]$ component of the $1\nu f_{7/2}$ orbital. Consequently, several near-degenerate configurations with the admixtures of $1f_{7/2}$ and $2p_{3/2}$  orbitals coexist at similar energies within these nuclei.  In the case of \nuc{S}{44}, two coexisting low-lying $0^+$ states have been observed~\cite{Grévy:2005S44isomer0+}, suggesting a spherical-deformed shape coexistence. The onset of deformed ground state in \nuc{S}{44} is supported by  the  quantum-number projected generator coordinate method (GCM) studies based on the Gogny force~\cite{Rodriguez:2011,Egido:2016}, and the collective Hamiltonian study based on  a relativistic energy density functional (EDF)~\cite{Li:2011edf}. These two theoretical studies reveal a general trend in shape evolution, i.e., the predominant shape transitions from a $\gamma$-soft, moderately deformed configuration with $\beta_2 \in [0.20, 0.30]$ to a strongly prolate shape with $\beta_2 \in [0.35, 0.45]$ as the angular momentum increases up to the $J=4$ in the ground-state band. This rotational band coexists with a strongly prolate-deformed $0^+_2$ state, characterized by $\beta_2 \in [0.35, 0.45]$. The specific deformation parameter $\beta_2$ of the dominant configuration depends on the details of the employed EDF. The two-proton knockout reaction from \nuc{Ar}{46} identified the $4_1^+$ state as an isomeric state~\cite{Santiago-Gonzalez:2011S444+}, which, in combination with shell-model calculations, was suggested to exhibit strong prolate deformation. A half-life measurement revealed a hindered $E2$ transition with $B(E2; 4^+_1 \to 2^+_1) = 0.61(19)$ W.u.~\cite{Parker:2017S44isomer4+}, supporting the interpretation of the $4^+_1$ state as a $K=4$ isomer. Shell-model studies further indicated that this state is predominantly characterized by the two-quasiparticle configuration $\nu 1/2^-[321] \otimes \nu 7/2^-[303]$~\cite{Utsuno:2014PRLS4443}.

The odd-mass neutron-rich sulfur isotope $\nuc{S}{43}$ exhibits a more complex low-energy structure than $\nuc{S}{44}$ due to the interplay between the single-particle motion of the unpaired neutron and the collective excitations of the $\nuc{S}{42}$ core. The mass measurements, combined with theoretical studies based on the shell model and relativistic mean-field (RMF) theory, suggested the coexistence of a prolate deformed 
ground state and an isomeric state in $\nuc{S}{43}$~\cite{Sarazin:2000PRL}. Subsequent $g$-factor measurements~\cite{Gaudefroy:2009PRL}, along with shell-model calculations and the collective Hamiltonian approach based on the Gogny force, determined the spin-parity of the isomeric state as $7/2^-_1$ at an excitation energy of 320.5(5) keV. These studies also established the intruder nature of the ground state with $K=1/2$~\cite{Gaudefroy:2009PRL}, while initially suggesting that the $7/2^-_1$ isomeric state is quasispherical.  However, later measurements of the spectroscopic quadrupole moment of the isomeric state yielded $Q_s(7/2^-_1)=23(3)$ $e$fm$^2$~\cite{Chevrier:2012PRL}, significantly larger than the expected value for a single-particle state. This observation indicates a strong collective nature of the isomeric state, which is further supported by shell-model calculations~\cite{Chevrier:2012PRL}. The structure of $\nuc{S}{43}$ was investigated using antisymmetrized molecular dynamics plus GCM (AMD+GCM), which predicted a prolate-deformed ground state, a triaxially deformed $7/2^-_1$ state, and an oblate-deformed excited band atop the $3/2^-_2$ state at low excitation energy~\cite{Kimura:2013AMD}. In contrast, a recent shell-model study suggested that the prolate ground-state band coexists with a triaxial band built on the $7/2^-_1$ isomer and an excited prolate structure associated with the $K^\pi=5/2^-$ deformed orbital~\cite{Chevrier:2014SMs43}.   Meanwhile, the angular-momentum-projected variation-after-projection (AMP+VAP) approach suggests that the ground state $3/2^-_1$ and the $7/2^-_1$ state are dominated by $K=1/2$ and $K=7/2$, respectively, classifying the $7/2^-_1$ state as a high-$K$ isomer~\cite{Utsuno:2014PRLS4443}.  Experimental lifetime measurements of excited states provide the first evidence of a doublet of $(3/2^-_2, 5/2^-_1)$ states. Together with shell-model and AMD calculations, these results suggest the possible existence of three coexisting bands built upon the $3/2^-_1$, $7/2^-_1$, and $3/2^-_2$ states~\cite{Mijatovic:2018PRL}. Furthermore, Coulomb excitation experiments have shown that the intraband $B(E2)$ values for the transitions within the ground-state band $3/2^-_1$ and the isomeric band $7/2^-_1$ are large and nearly equal~\cite{Longfellow:2020PRL}, which was also found in the study of valence-space in-medium similarity renormalization group~\cite{Yuan:2024VSIMSRG}.

Over the past decades, covariant density functional theory (CDFT) has achieved remarkable success in various areas of nuclear physics~\cite{Vretenar:2005PR,Meng:2006,Meng:2016Book}. A key advantage of CDFT is that Lorentz invariance imposes strict constraints on the number of parameters in the EDF. Moreover, the relativistic framework naturally accounts for the spin-orbit interaction, while time-odd fields are incorporated without introducing additional free parameters. This characteristic is particularly crucial for accurately describing odd-mass nuclei and rotating systems. To restore the missing quantum numbers, including particle numbers and angular momentum, in the solution of CDFT and to consider the shape-mixing effect,  the multi-reference covariant density functional theory (MR-CDFT) has been developed~\cite{Yao:2009,Yao:2010,Niksic:2011ppnp} and successfully applied to study low-lying spectra in even-even nuclei with either triaxial or octupole shapes~\cite{Yao:2014_76Kr,Yao:2015_Octupole,Zhou:2023JMPE}. The MR-CDFT has also been applied to the studies of neutrinoless double-beta decay~\cite{Yao:2021PPNP} and the low-lying states of hypernuclei~\cite{Mei:2016PRC,Xia:2019,Xia:2023}.

Recently, the MR-CDFT was successfully extended to low-lying states of odd-mass nuclei~\cite{Zhou:2024}. In this paper, we extend the MR-CDFT further with the inclusion of both $K$-mixing and shape-mixing effects and present the first application of this extended MR-CDFT to the low-lying states of \nuc{S}{43}.  This novel implementation allows us to identify the dominant effect responsible for the isomeric state, offering new insights into the interplay of deformation and $K$-forbiddenness in odd-mass nuclei. The coexistence of rotational bands built on top of three different configurations is also investigated. 
  
The article is arranged as follows. In Sec.\ref{sec:framework}, we present the framework of MR-CDFT for odd-mass nuclei.  The results of calculations for \nuc{S}{43} are discussed in Sec.\ref{sec:results}. The conclusion of this study is summarized in Sec.~\ref{sec:summary}.

  \section{The MR-CDFT for odd-mass nuclei }
 \label{sec:framework} 
   
The MR-CDFT theory for the low-lying states of odd-mass nuclei has been introduced in detail in Ref.~\cite{Zhou:2024}. Here, we only present a brief description of this theory, in which the wave functions of low-lying states are constructed as a mixing of configurations with different deformation parameter $\mathbf{q}$ and quantum number $K$,  
\begin{equation}
\label{eq:gcmwf}
\vert \Psi^{J\pi}_\alpha\rangle
=\sum_{c} f^{J\alpha \pi}_{c}  \ket{NZ J\pi; c}.
\end{equation} 
Here, $c$ is a collective label for $(K,\mathbf{q})$, and $\alpha$ distinguishes states with the same angular momentum $J$. The basis function with quantum numbers ($NZJ\pi$) is given by
\begin{equation}
\label{eq:basis}
\ket{NZ J\pi; c} 
=  \hat P^J_{MK} \hat P^N\hat P^Z \ket{\Phi^{\rm (OA)}_\kappa(\mathbf{q})},
\end{equation}
where $\hat P^{J}_{MK}$ and $\hat{P}^{N(Z)}$  are projection operators that select components with the angular momentum $J$, and neutron(proton) number $N(Z)$~\cite{Ring:1980}. 

 The mean-field configurations $\ket{\Phi^{\rm (OA)}_\kappa(\mathbf{q})}$ for odd-mass nuclei are chosen as one-quasi-particle (1qp) states,
\begin{eqnarray}
\label{eq:odd-mass-wfs}
 \ket{\Phi^{\rm (OA)}_\kappa(\mathbf{q})}  =\alpha^\dagger_\kappa \ket{\Phi_{(\kappa)}(\mathbf{q})},
\end{eqnarray} 
where $\ket{\Phi_{(\kappa)}}$ denotes a quasiparticle vacuum state with even number parity obtained through the false quantum vacuum(FQV) scheme~\cite{Zhou:2024} in the single reference (SR)-CDFT calculation stating from a relativistic EDF~\cite{Zhao:2010,Meng:2016Book}. The quasiparticle creation operator $\alpha^\dagger$ switches the number parity to be odd. The index $\kappa$ distinguishes different quasiparticle states. In the present study, axial symmetry is assumed. In this case,  each configuration is labeled with the quantum numbers $K^\pi$, which  are determined by the Nilsson quantum number $\Omega^\pi$ of the blocked orbital, i.e.,  $K^\pi=\Omega^\pi$.

The weight function $f^{J\alpha \pi}_{c}$ in Eq.~(\ref{eq:gcmwf}) is determined by the variational principles which lead to the following Hill-Wheeler-Griffin (HWG) equation~\cite{Hill:1953,Ring:1980},
\begin{eqnarray}
\label{eq:HWG}
\sum_{c'}
\Bigg[\mathscr{H}^{NZJ\pi}_{cc'}
-E_\alpha^{J\pi }\mathscr{N}^{NZJ\pi }_{cc'} \Bigg]
f^{J\alpha \pi}_{c'}=0,
\end{eqnarray}
where the Hamiltonian kernel  and norm kernel are defined by
\begin{equation}
\label{eq:kernel}
 \mathscr{O}^{NZJ\pi}_{cc'}
 =\bra{NZ J\pi; c}  \hat O \ket{NZ J\pi; c'},
\end{equation}
with the operator $\hat O$ representing $\hat H$ and $1$, respectively. In the present study based on a covariant EDF,  the mixed-density prescription is employed in the evaluation of the Hamiltonian kernel. Details on the calculation of the kernels in Eq.~(\ref{eq:kernel}) can be found in Ref.~\cite{Zhou:2024}.

The HWG equation (\ref{eq:HWG}) for a given set of quantum numbers $(NZJ\pi)$ is solved in the standard way as discussed in Refs.~\cite{Ring:1980,Yao:2010}. This is done by first diagonalizing the norm kernel $\mathscr{N}^{NZJ\pi }_{cc'}$. A new set of basis functions is then constructed using the eigenfunctions of the norm kernel with eigenvalues larger than a pre-chosen cutoff value, which removes possible redundancy in the original basis.  The Hamiltonian is diagonalized on this new basis. In this way, the energies $E_\alpha^{J\pi}$ and
the mixing weights $f^{J\alpha\pi}_{c}$ of the nuclear states $\vert \Psi^{J\pi}_\alpha\rangle$ can be obtained. Since the basis functions $\ket{NZ J\pi; c}$ are nonorthogonal to each other, one usually introduces the collective wave function $g^{J\pi}_\alpha(K,\mathbf{q})$ as below
\begin{equation}
\label{eq:coll_wf}
g^{J\pi}_\alpha(K,\mathbf{q})=\sum_{c'} (\mathscr{N}^{1/2})^{NZJ\pi}_{c,c'} f^{J\alpha\pi}_{c'},
 \end{equation}
 which fulfills the normalization condition. The distribution of $g^{J\pi}_\alpha(K,\mathbf{q})$ over $K$ and $\mathbf{q}$ reflects the contribution of each basis function to the nuclear state $\vert \Psi^{J\pi}_\alpha\rangle$. With the mixing weight $f^{J\alpha\pi}_{c}$, it is straightforward to determine the observables of nuclear low-lying states, including electric quadrupole moment $Q_s$, magnetic dipole moment $\mu$, as well as the $E2$ and $M1$ transition strengths. 
 The strength of the $E\lambda (M\lambda)$ transition from the initial state $\ket{\Psi^{J_i\pi_i}_{\alpha_i}}$ to the final state $\ket{\Psi^{J_f\pi_f}_{\alpha_f}}$ is determined by 
\begin{eqnarray}
&& B(T\lambda,J_i\alpha_i\pi_i\rightarrow J_f\alpha_f\pi_f) = \cfrac{1}{2J_i+1} \nonumber\\
&&\times\left| \sum_{c_f, c_i} f^{J_i\alpha_i \pi_i}_{c_i}f^{J_f\alpha_f \pi_f}_{c_f} 
    \langle NZJ_f\pi_f,c_f||\hat T_\lambda ||NZJ_i\pi_i,c_i\rangle
    \right|^2,
\end{eqnarray} 
where the configuration-dependent reduced matrix element is simplified as follows, 
\begin{eqnarray}
\label{eq:reduced_matrix_element}
    &&\langle NZ J_f\pi_f; c_f ||\hat  T_\lambda|| NZ J_i\pi_i; c_i\rangle\nonumber\\
    =&& \delta_{\pi_f\pi_i,(-1)^\lambda}(-1)^{J_f-K_f}
    \cfrac{\hat{J}_i^2\hat{J}_f^2}{8\pi^2} \sum_{\nu M} \left(
 \begin{array}{ccc}
J_f&\lambda&J_i\\
-K_f &\nu&M\\
\end{array}
    \right) \nonumber\\
   &&\times  \int d\Omega D_{MK_i}^{J_i\ast}(\Omega)
\bra{\Phi^{\rm (OA)}_{\kappa_f}(\mathbf{q}_f)}
    \hat T_{\lambda\nu}
  \hat R(\Omega) \hat P^Z\hat P^N\hat P^{\pi_i}
   \ket{\Phi^{\rm (OA)}_{\kappa_i}(\mathbf{q}_i)},\nonumber \\
\end{eqnarray}
where $\hat T_{\lambda\nu}$ represents either an electric or a magnetic multipole operator and $\hat J=\sqrt{2J+1}$.
The detailed formulas can be found in Ref.~\cite{Zhou:2024}.

 \section{Results and discussion}
 \label{sec:results}

In the calculation of the mean-field configurations, the Dirac spinors for single nucleons are solved using a harmonic oscillator basis with a major shell number of $N_{\rm sh} = 10$. Pairing correlations between nucleons are treated within the BCS approximation using a density-independent $\delta$ force with a smooth cutoff~\cite{Krieger:1990pks,Yao:2009}. The PC-PK1 parameterization is employed for the relativistic EDF~\cite{Zhao:2010}. In the calculation of the projected kernels, the number of mesh points in the interval $[0, \pi]$ for the rotation angle $\beta$ and the gauge angle $\varphi$ are chosen as $N_\beta = 12$ and $N_{\varphi} = 7$, respectively. These values are found to be sufficient to achieve convergent results for \nuc{S}{43}.

\begin{figure}[tb]
 \centering
\includegraphics[width=8cm]{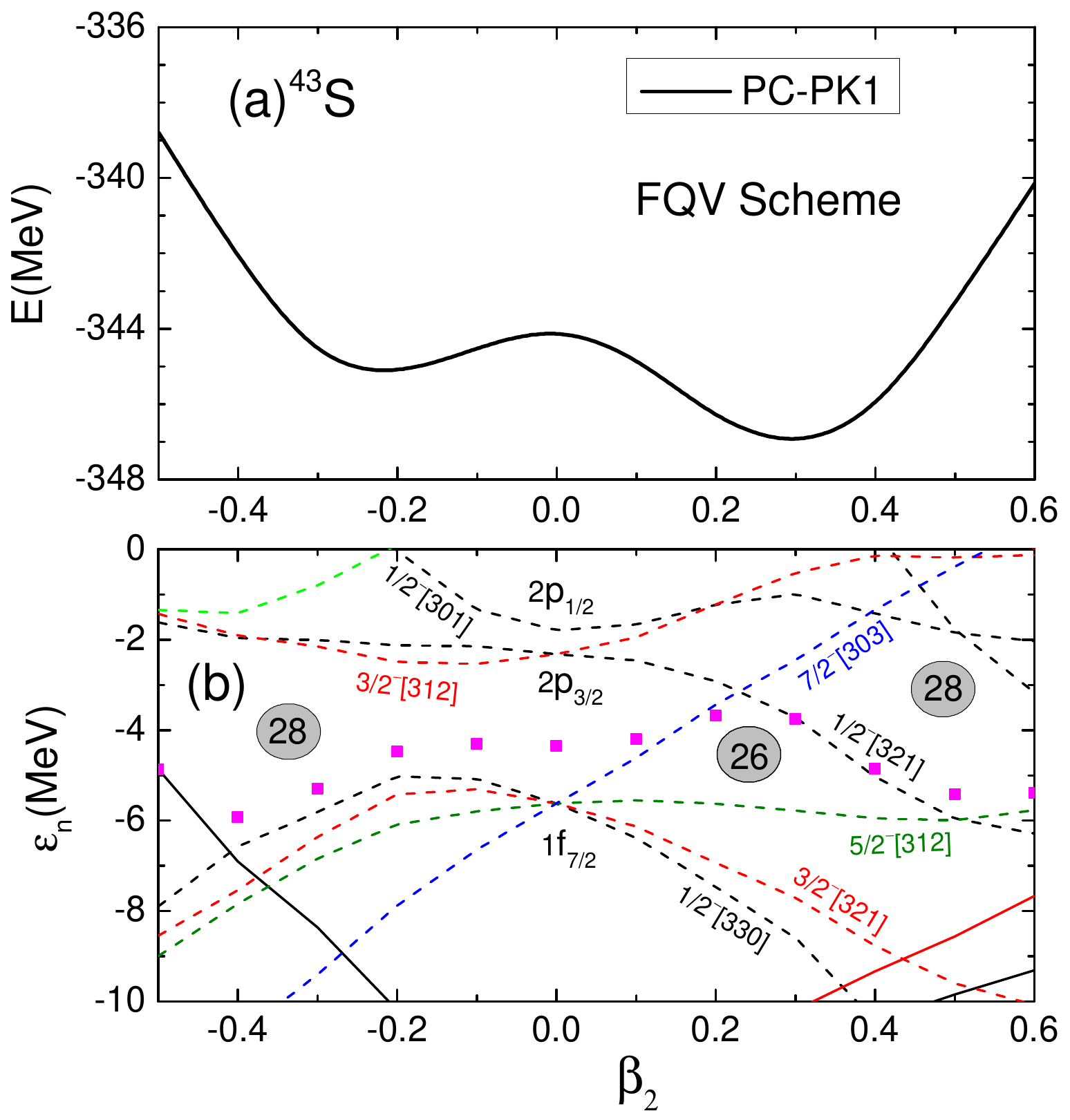}
\caption{\label{fig:S43_MFE_SG} (Color online) (a) Total energy of the mean-field states for \nuc{S}{43} from the SR-CDFT calculation based on the FQV scheme~\cite{Zhou:2024}, as a function of the quadrupole deformation parameter $\beta_2$.
(b) Nilsson diagram for neutrons in \nuc{S}{43}, with the Fermi energies indicated by pink squares. The numbers around the lines are the Nilsson quantum numbers $\Omega^\pi[Nn_z m_\ell]$~\cite{Ring:1980}. }
 \end{figure}

Figure~\ref{fig:S43_MFE_SG}(a) presents the energies of mean-field states $\ket{\Phi_{(\kappa)}(\mathbf{q})}$ from the SR-CDFT calculation based on the FQV scheme~\cite{Zhou:2024}  as a function of the quadrupole deformation parameter $\beta_2$. A pronounced energy minimum appears on the prolate side with $\beta_2\simeq 0.3$, with a second minimum  on the oblate side with $\beta_2\simeq-0.2$, suggesting that \nuc{S}{43} may exhibit coexisting prolate and oblate shapes in its low-lying states. This phenomenon can be understood from the Nilsson diagram of neutrons in \nuc{S}{43}, as shown in Fig.~\ref{fig:S43_MFE_SG}(b). One can see that the downward $\nu1/2^-[321]$ component of the $2\nu p_{3/2}$ orbital crosses with the upward $\nu7/2^-[303]$ component of the $1\nu f_{7/2}$ orbital at $\beta_2\simeq 0.22$ around the Fermi energy. This crossing leads to the population of valence neutrons from $\nu7/2^-[303]$ to $\nu1/2^-[321]$. As a result, a large $N=28$ shell gap shows up around $\beta_2=0.5$. This result is consistent with the result of AMD+GCM calculation \cite{Kimura:2013AMD}. On the oblate side, the $N=28$ shell gap increases with the $|\beta_2|$.

\begin{figure}[tb]
 \centering
\includegraphics[width=8.4cm]{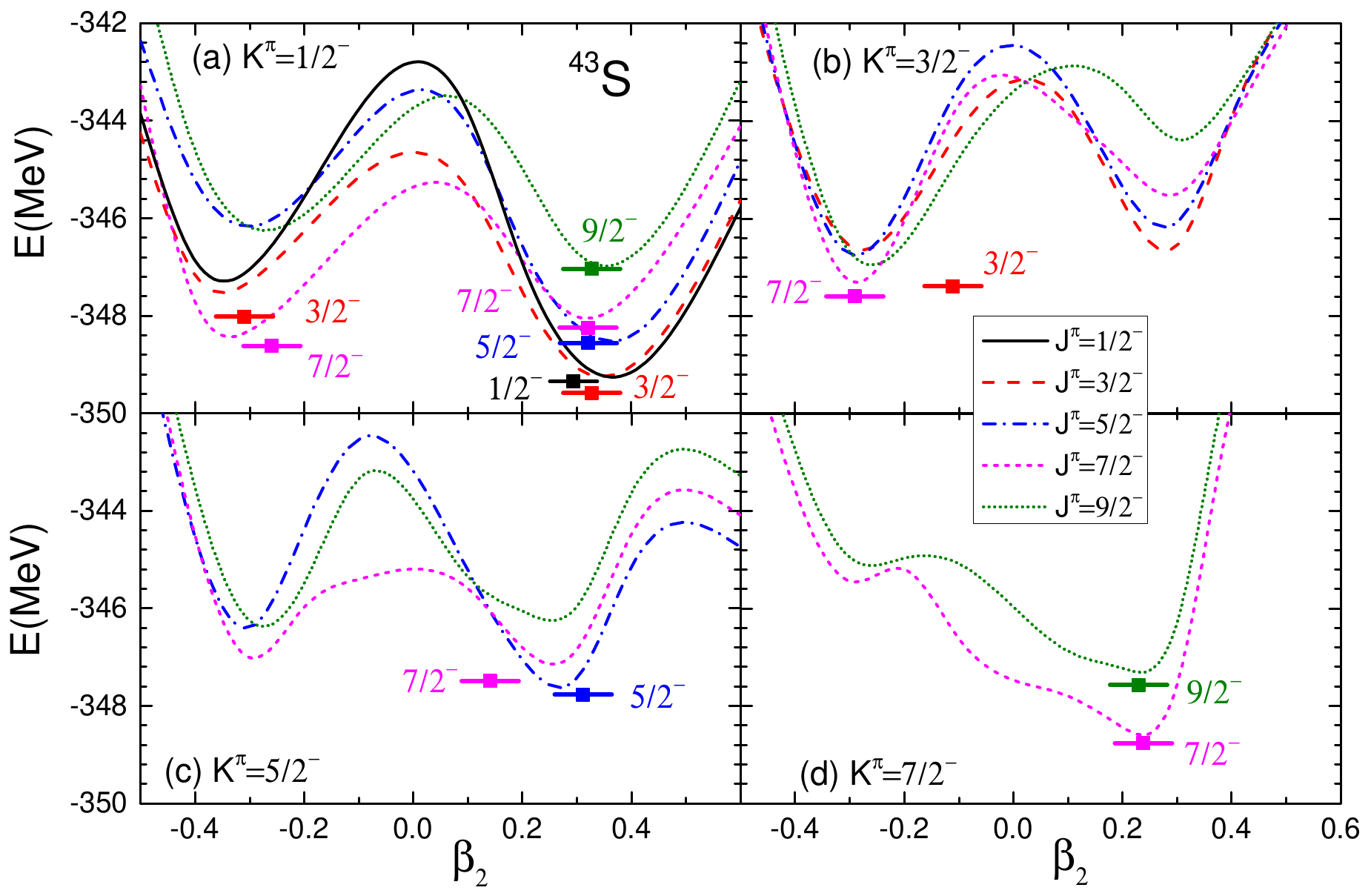}
\caption{(Color online) 
Energies of symmetry-conserving 1qp states with different angular momenta in \nuc{S}{43} for (a) $K^\pi = 1/2^-$, (b) $K^\pi = 3/2^-$, (c) $K^\pi = 5/2^-$, and (d) $K^\pi = 7/2^-$, as functions of the quadrupole deformation parameter $\beta_2$. The discrete low-lying states obtained by mixing  configurations of different shapes within the GCM framework are shown at their mean quadrupole deformation $\bar{\beta}_2^{J\pi\alpha}$, as defined in Eq.~(\ref{eq:mean_beta2}).
 }
 \label{fig:AMP_energy} 
 \end{figure}

The wave functions of the mean-field states $\ket{\Phi_{(\kappa)}(q)}$ in Fig.~\ref{fig:S43_MFE_SG}(a) do not preserve the  particle numbers or total angular momentum. By applying quantum-number projection operators onto the 1qp configurations with $K^\pi=1/2^-, 3/2^-, 5/2^-$, and $7/2^-$, c.f. Eqs.(\ref{eq:basis}) and (\ref{eq:odd-mass-wfs}), one obtains the energies of symmetry-conserved 1qp states in \nuc{S}{43}, as displayed in Fig.~\ref{fig:AMP_energy}.  The low-lying states from the shape-mixing calculation are  plotted at their mean quadrupole deformation  $\bar \beta_2^{J\pi\alpha}$, which is defined as
\beq
\label{eq:mean_beta2}
 \bar{\beta}_2^{J\pi\alpha}(K^\pi)
=\sum_{\beta_2} |g^{J\pi}_\alpha(K, \beta_2)|^2\beta_2.
\eeq 
The energies of symmetry-conserving states with angular momentum $J$ increasing from $1/2$ to $9/2$, projected out from the configurations with $K^\pi=1/2^-$ as a function of the quadrupole deformation $\beta_2$ are displayed in Fig.~\ref{fig:AMP_energy}(a). Similar to the energy curve of the mean-field calculation in Fig.~\ref{fig:S43_MFE_SG}(a), all the projected energy curves present two energy minima on the prolate and oblate side, respectively, with $|\beta_2|\simeq 0.3$. It is interesting to note the change of the energy ordering of states with different angular momenta as a function of the deformation from Fig.~\ref{fig:AMP_energy}(a). For the configurations with $\beta_2>0.4$, the energies of projected states with $\Delta J=1$ follow the ordering  of $(1/2^-, 3/2^-,  5/2^-, 7/2^-, 9/2^-)$, which are consistent with the {\em strong} coupling limit of the particle-rotor model (PRM)~\cite{Ring:1980}. In contrast, for the weakly deformed configurations with $\beta_2 < 0.3$, the $1/2^-$ state rises rapidly from the bottom as $\beta_2$ decreases toward zero. When considering only the configurations of the oblate energy minima, the energy ordering becomes $(7/2^-, 3/2^-,  1/2^-, 9/2^-, 5/2^-)$. After mixing the configurations with different shapes, but with the same quantum numbers $K^\pi=1/2^-$, one obtains the yrast states with the energy ordering of $(3/2^-, 1/2^-,  5/2^-, 7/2^-)$ mainly located around $\beta_2=0.3$.

  \begin{figure*}[]
 \centering
\includegraphics[width=15cm]{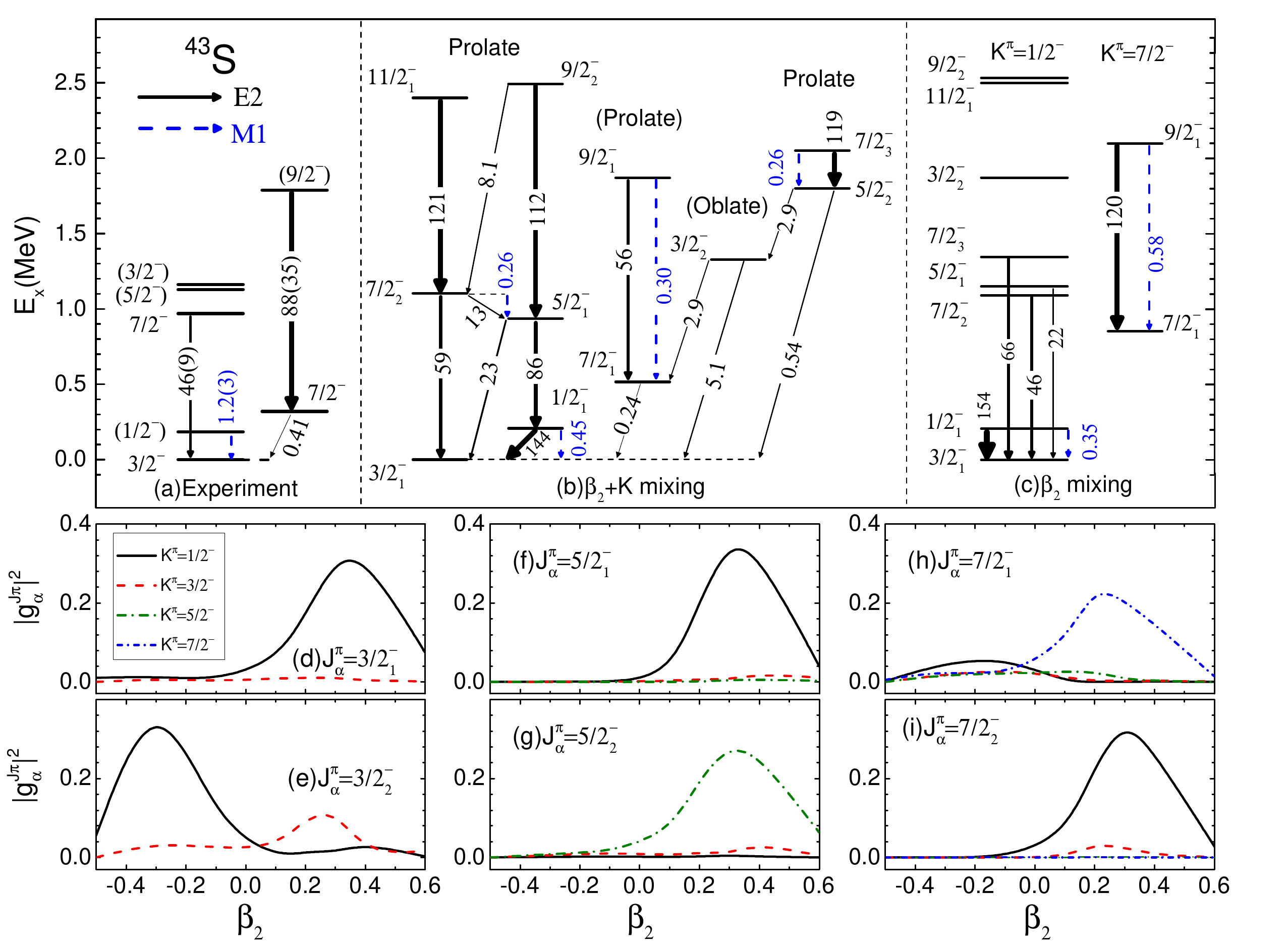}
\caption{\label{fig:S43_level} (Color online) 
Energy spectra of low-lying states in \nuc{S}{43} from MR-CDFT calculations. Panel (b) shows the results including configuration mixing with different shapes and $K$ values, while panel (c) shows the results without $K$ mixing. Experimental data~\cite{Gaudefroy:2009PRL,Longfellow:2020PRL} are displayed in panel (a). Black and blue arrows indicate the $B(E2)$ values (in units of $e^2$fm$^4$) and $B(M1)$ values (in units of $\mu_N^2$), respectively. The states dominated by a prolate (oblate) configuration, with a small admixture of oblate (prolate) components, are labeled as ``(Prolate)'' and ``(Oblate)'', respectively.  The distributions of the collective wave functions $|g^{J\pi}_\alpha(K,\beta_2)|^2$ for the first two states with $J = 3/2$, $5/2$, and $7/2$ are shown in panels (d, e), (f, g), and (h, i), respectively.   
}
 \end{figure*}

  Figure~\ref{fig:AMP_energy} (b), (c) and (d) display the energies of symmetry-conserving states with $J\geq K$, projected out from the configurations with $K^\pi=3/2^-, 5/2^-$, and $7/2^-$, respectively. Besides, the two lowest energy states from the shape-mixing calculation are given. One can see that the energy level $7/2^-$ dominated by the prolate configuration with $K^\pi=7/2^-$ is the yrast state among all the $J^\pi=7/2^-$ states. Actually, all the states with the same quantum numbers $J^\pi$ but different $K$ values might be mixed. Those from the MR-CDFT calculation with both $K$-mixing and shape-mixing effects are plotted in Fig.~\ref{fig:S43_level}(b). The distributions of the collective wave functions for the first two $3/2^-, 5/2^-$, and $7/2^-$ states are shown in the bottom panels of Fig.~\ref{fig:S43_level}.
  It is seen that the main features of the measured low-lying states of \nuc{S}{43}, cf. Fig.~\ref{fig:S43_level}(a),  are reasonably well reproduced only in the MR-CDFT calculations that incorporate  the mixing of both $K$ values and  shapes with different $\beta_2$ values. 
  
  Here, we summarize the main findings from the MR-CDFT study based on  Fig.~\ref{fig:S43_level}, as follows.  The ground state ($3/2^-_1$) of \nuc{S}{43} is predominantly characterized by a prolate-deformed configuration, $\nu 1/2^-[321]$, with $K^\pi = 1/2^-$. In contrast, the $3/2^-_2$ state is dominated by an oblate configuration with $\beta_2 \simeq -0.3$ and $K^\pi = 1/2^-$, with a small admixture of a prolate configuration having $K^\pi = 3/2^-$. We note that in the AMD+GCM calculation~\cite{Kimura:2013AMD}, the $3/2^-_2$ state is dominated by an oblate configuration with $K^\pi = 3/2^-$, followed by the $5/2^-_2$ and $7/2^-_3$ states, forming an oblate rotational band. As shown in Fig.\ref{fig:S43_level}(b), we predict a strong $E2$ transition between the $5/2^-_2$ and $7/2^-_3$ states, whereas the $E2$ transition from the $5/2^-_2$ to the $3/2^-_2$ state is much weaker. This behavior can be understood from Fig.\ref{fig:S43_level}(g), which shows that the $5/2^-_2$ state is dominated by a prolate-deformed configuration with $K^\pi = 5/2^-$, in contrast to the oblate structure of the $3/2^-_2$ state.
 
  The first excited state of \nuc{S}{43}, $1/2^-_1$, shares a similar predominant configuration with the ground state and exhibits a strong $E2$ transition to the ground state. The MR-CDFT (labeled with ``$\beta_2+$K mixing'') predicts $B(E2; 1/2^-_1 \to 3/2^-_1) = 144$ $e^2$fm$^4$, as well as a sizable $M1$ transition, with $B(M1; 1/2^-_1 \to 3/2^-_1) = 0.45$ $\mu_N^2$, as shown in  Fig.~\ref{fig:S43_level}(b). In the AMD+GCM~\cite{Kimura:2013AMD}, these numbers are 180 $e^2$fm$^4$ and 0.34 $\mu_N^2$, respectively. The value of $B(E2; 1/2^-_1 \to 3/2^-_1)$ has not yet been measured, but experimental data are available for $B(M1; 1/2^-_1 \to 3/2^-_1) = 1.2(3)$ $\mu_N^2$\cite{Mijatovic:2018PRL}, which is underestimated by both methods. We find that this underestimation is due to the extension of the wave function of the $3/2^-_1$ state into the region of large prolate deformation with $\beta_2 > 0.4$, as shown in Fig.\ref{fig:S43_level}(d). Moreover, we observe two $\Delta J = 2$ rotational bands, namely ($3/2^-_1$, $7/2^-_2$, $11/2^-_1$) and ($1/2^-_1$, $5/2^-_1$, $9/2^-_2$), which are connected by strong $E2$ transitions. The predicted $B(E2; 7/2^-_2 \to 3/2^-_1) = 59$ $e^2$fm$^4$ is slightly larger than the data 46(9) $e^2$fm$^4$~\cite{Longfellow:2020PRL}. All these states are predominantly characterized by the prolate-deformed configuration $\nu 1/2^-[321]$, originating from the spherical $\nu p_{3/2}$ orbital, with deformation $\beta_2 \simeq 0.3$ and $K^\pi = 1/2^-$, as illustrated in Fig.\ref{fig:S43_MFE_SG}. The energy ordering of the ($3/2^-_1$, $7/2^-_2$, $11/2^-_1$) sequence follows the {\em weak} coupling limit of the particle-rotor model~\cite{Ring:1980}, which predicts a parabolic energy pattern with a minimum at $J \simeq j = 3/2$.
    
 Figure~\ref{fig:S43_level}(h) shows that the $7/2^-_1$ state is dominated by the prolate deformed configuration  $\nu 7/2^-[303]$ with $\beta_2\simeq 0.22$ and $K^\pi = 7/2^-$. Consequently, the decay of the $7/2^-_1$ state to the ground state ($3/2^-$) is strongly quenched due to $\Delta K = 3$. Quantitatively, the predicted $B(E2; 7/2^-_1 \to 3/2^-_1) = 0.24$ $e^2$fm$^4$, compared to the data 0.41 $e^2$fm$^4$~\cite{Gaudefroy:2009PRL}. As a result, the $7/2^-_1$ state is classified as a high-$K$ isomer state, consistent with previous studies~\cite{Chevrier:2014SMs43,Utsuno:2014PRLS4443}.  A comparison between Figs.\ref{fig:S43_level}(b) and (c) reveals that the excitation energy of the $7/2^-_1$ state decreases significantly and becomes closer to the experimental value after incorporating the $K$-mixing effect. As shown in Fig.\ref{fig:S43_level}(h), the isomeric $7/2^-_1$ state also contains a small admixture of an oblate configuration with $K^\pi = 1/2^-$. The remaining discrepancy is expected to be further reduced by including triaxial deformation effects, as suggested by the AMD+GCM calculation~\cite{Kimura:2013AMD}.

\begin{table}[tb]
\centering
    \tabcolsep=2pt
    \caption{The spectroscopic quadrupole moments $Q_s(J_\alpha^\pi)$ and magnetic dipole moments $\mu(J_\alpha^\pi)$ for $^{43}$S based on MR-CDFT calculations, in comparison with the prediction of
 AMD calculations~\cite{Kimura:2013AMD} and data~\cite{Chevrier:2012PRL}.}
    \begin{tabular}{cccrrrcrr}
      \hline \hline 
       & \multicolumn{3}{c}{$Q_s ~$($e$~fm$^2$)} 
       &&&& \multicolumn{1}{c}{$\mu(\mu_N)$}\\   
       \hline
 $J_\alpha^\pi$ &  & Exp. & MR-CDFT &  AMD& & Exp.  & MR-CDFT  &AMD  \\  
 \hline 
 $1/2^-_1$ &  & - &    -    &    -    &  & -  &  $0.48$  &  0.71    \\  
 $3/2^-_1$ &  & - & $-13.0$ & $-13.2$ &  & -  &  $-0.76$ & $-0.60 $ \\  
 $5/2^-_1$ &  & - & $-17.4$ & $-20.1$ &  & -  &  $1.49$  & 1.45     \\  
 $7/2^-_2$ &  & - & $-18.9$ & $-22.0$ &  & -  &  $-0.12$ & $-0.24$  \\  
 $9/2^-_2$ &  & - & $-24.1$ & $-21.4$ &  & -  &  $2.20$  & 1.26     \\ 
 $3/2^-_2$ &  & - & $ 9.2$  & $12.1$  &  & -  &  $-0.54$ & $-0.82$  \\ 
 $5/2^-_2$ &  & - & $ 18.1$ & $-7.0$  &  & -  &  $-0.85$ & 0.14     \\ 
 \hline 
 $7/2^-_1$ &  &23(3)& $20.9$ & $26.1$ &  &  $-1.110(14)$ & $-0.93$ &  $-1.08$ \\  
 $9/2^-_1$ &  & -   & $4.9$  & $7.3$  &  &     -         & $0.08$  &  $-0.19$ \\  
\hline\hline 
    \end{tabular}
    \label{tab:Qs_mu_S43}
\end{table}

Table~\ref{tab:Qs_mu_S43} lists the spectroscopic quadrupole moments $Q_s$ and magnetic dipole moments $\mu$ for \nuc{S}{43} obtained from the MR-CDFT calculations with both shape-mixing and $K^\pi$ mixing effects, in comparison with the results of AMD+GCM calculations~\cite{Kimura:2013AMD} and available data for the isomer state with $Q_s (7/2^-_1)=23(3)$ $e$ fm$^2$ and $\mu(7/2^-_1)=-1.110(14) \mu^2_N$. Overall, the results from MR-CDFT and AMD+GCM are similar, with both models reasonably reproducing the experimental data. Quantitatively, the $Q_s$ values of states in the ground-state band predicted by MR-CDFT are slightly smaller than those from AMD+GCM. Notably, the magnetic moments of the $1/2^-_1$ and $7/2^-_2$ states obtained with MR-CDFT are only about half the values predicted by AMD+GCM. A significant difference arises in the prediction for the $5/2^-_2$ state: while the AMD+GCM calculation indicates that this state is dominated by an oblate configuration, the MR-CDFT result suggests a prolate deformation, consistent with the latest shell-model calculations~\cite{Chevrier:2014SMs43}. Further experimental and theoretical benchmarks are needed to ultimately validate these findings.

 \section{Summary}
 \label{sec:summary}
In this work, we have extended the multireference covariant density functional theory (MR-CDFT) by incorporating particle-number and angular-momentum projections to describe the low-lying states of \nuc{S}{43}, including configuration mixing with different intrinsic shapes and $K$ quantum numbers within the framework of the generator coordinate method (GCM). The available experimental data on energy spectra, electric quadrupole and magnetic dipole transition strengths, as well as electric quadrupole and magnetic dipole moments, are reasonably well reproduced, demonstrating the capability of MR-CDFT to describe the low-energy structure of odd-mass nuclei around $N=28$ where shape coexistence and $K$-mixing play a crucial role.
Our calculations reveal a pair of prolate rotational bands built on the ground-state configuration $\nu1/2^-[321]$ with $K^\pi = 1/2^-$,  belonging to the weak-coupling limit of the particle rotor model, and another rotational band built on the configuration $\nu7/2^-[303]$ with $K^\pi = 7/2^-$, corresponding to the isomeric $7/2^-_1$ state. These findings are generally consistent with those from previous studies using antisymmetrized molecular dynamics plus GCM (AMD+GCM) and shell-model calculations, supporting the picture of $N=28$ shell erosion.

Additionally, we identify a $3/2^-_2$ state dominated by an oblate configuration with $K^\pi = 1/2^-$, where the valence neutron occupies the $\nu1/2^-[330]$ orbital, and a $5/2^-_2$ state dominated by a prolate configuration with $K^\pi = 5/2^-$. This latter result differs from AMD+GCM predictions, which favor an oblate shape for the $5/2^-_2$ state. To further clarify the nature of this state, lifetime measurements are necessary.
Finally, we note that the excitation energy of the isomeric $7/2^-_1$ state is still somewhat overestimated, likely due to the omission of triaxial deformation effects. An extension of the MR-CDFT framework to incorporate triaxiality in odd-mass nuclei is currently in progress.

\section*{acknowledgments} 
This work was partially supported by the National Natural Science Foundation of China (Nos. 12465020, 12005802, 12375119, and 12141501), the Guangdong Basic and Applied Basic Research Foundation (2023A1515010936), the Fundamental Research Funds for the Central Universities, Sun Yat-sen University, and the Deutsche Forschungsgemeinschaft (DFG, German Research Foundation) under Germany’s Excellence Strategy – EXC-2094-390783311, ORIGINS.

 \bibliographystyle{apsrev4-1} 
 
%

\end{document}